# Absence of a Competition between Magnetism and Superconductivity in the Layered Nickel Borocarbides: Common Correlation $T_c$ with Crystal Chemical Parameters for Magnetic and Nonmagnetic Compounds

**L.M. Volkova[1], S.A. Polyshchuk[1], S.A. Magarill[2], and F. E. Herbeck[3]**

[1] *Institute of Chemistry, Far Eastern Branch of Russian Academy of Science, 690022 Vladivostok, Russia*

[2] *Institute of Inorganic Chemistry, Sib. Branch of Russian Academy of Science, 630090 Novosibirsk, Russia*

[3] *Institute of Automation and Control Processes, Far Eastern Branch of Russian Academy of Science, 690041 Vladivostok, Russia.*

In the work the analysis of crystal chemical researches of nickel borocarbides $RNi_2B_2C$ (R = rare earth) is given. The reasons of representation of dependence of superconducting transition temperature ($T_c$) from crystal chemical parameters by two separate curves for magnetic and nonmagnetic R are considered. Common for all R dependences of $T_c$ from crystal chemical parameters similar existing in layered quasi-two-dimensional systems (HTSC cuprates and diborides) are established. The absence of influence on borocarbides $T_c$ of magnetic properties R is determined. On the basis of the found correlations the radii of a number of rare earths are precised and $T_c$ of compounds at various substitutions R are calculated.

## 1. INTRODUCTION

The superconducting, magnetic and structural characteristics of layered nickel borocarbides $RNi_2B_2C$ (R = rare earth) are comprehensively investigated. It was established, that borocarbides with the nonmagnetic rare earth ions Sc, Y and Lu have the highest $T_c$ (13.5 K - 16.6 K) as compared to the magnetic rare earth ions such as Dy, Ho, Er and Tm (6 K -11 K). According to the generally accepted opinion, the magnetic properties rare earth ions surpress partially or completely superconducting properties in $RNi_2B_2C$ compounds [1,2]. It is necessary to underline, that this conclusion occures from difficulties in installation for magnetic and nonmagnetic rare earth ions of common dependencies of $T_c$ from crystal chemical parameters of these compounds. Separately for nonmagnetic ions R it was possible to fix the dependence of $T_c$ from the following parameters: a size R ions, Ni-Ni distances [3] and the ratio c/a lattice parameters [4].

There is also other point of view that in $RNi_2B_2C$ compounds, the interaction between the magnetic moment of the R ion and conduction electrons are not strong enough to produce pair breaking effects and the destruction of superconductivity [5]. This conclusion is grounded on the coexistence of superconductivity and magnetic order in $RNi_2B_2C$ (R = Tm, Er, Ho and Dy) [6-13]. Moreover, it is supposed in [5] that the R ions do play a role in the onset of superconductivity and the size of the ion also has an effect on superconductivity in these compounds. This supposition was doubtless confirmed by the fact of $T_c$ decrease to complete suppression of superconductivity in $R_{1-x}R'_xNi_2B_2C$ by introduction both magnetic and nonmagnetic impurities R' of rare-earth ions, which radius is more than ionic radius R (in $Ho_{1-x}R'_xNi_2B_2C$ R = La [14], Dy [15]; in $Y_{1-x}R'_xNi_2B_2C$ R = Pr [16], Tb [17, 18] Ho [14, 19] and in $Lu_{1-x}Ho_xNi_2B_2C$ [14]) Moreover, the effect of suppression of superconductivity clearly correlates with increase of difference of the ionic radii (system of ionic radii of Pauling) R of "host" and impurity R'.

The data about common for magnetic and nonmagnetic rare earths dependence $T_c$ from crystal chemical parameters of $RNi_2B_2C$ are well grounded confirmation of absence of a competition between magnetism and superconductivity in nickel borocarbides. The similar features of structure of the layered borocarbides with HTSC cuprates and diborides admit possibility of existence of common dependence for borocarbides, as was found by us for HTSC-cuprates [23-25]. Thus, the purpose of work is the search of this dependence on the basis of more careful analysis of experimental data on structures and superconducting properties of nickel borocarbides, represented in the literature.

## 2.THE ANALYSIS OF CRYSTAL CHEMICAL DEPENDENCIES IN $RNi_2B_2C$

We have used the experimental data (structural parameters and $T_c$) of 50 compounds $RNi_2B_2C$ (R = Sc, Y, La, Pr, Nd, Gd, Tb, Dy, Ho, Er, Tm, Yb Lu, Ce, $Y_{1-x}Lu_x$ and $Tb_{1-x}Y_x$) [4, 6, 11, 18, 26-33] (Tabl. 1), excepting Yb borocarbide, guessing, that the absence of superconducting properties in it is bound with scanty impurity of divalent ytterbium. For borocarbides with several sorts of rare earths ions or containing R with the valence, unequal 3 ($R'_{1-x}R'_xNi_2B_2C$ and $CeNi_2B_2C$) instead of radius of a rare-earth ion effective radius $\boldsymbol{r^{eff}}$ was utilized, which is the generalized value describing a size and a charge of R-ions, and dispersion of these parameters:

$$r^{eff} = S\overline{r_R(Z_R/3)} \qquad (1)$$

where $\overline{r_R(Z_R/3)}$ is the value describing an averaged size and a charge of ions R:

$$\overline{r_R(Z_R/3)} = m_1 r_{R_1}(Z_{R_1}/3) + \ldots m_n r_{R_n}(Z_{R_n}/3) \qquad (2)$$

**Table I.** Structural Parameters and $T_c$ of Borocarbides $RNi_2B_2C$ Used for Calculation

| N | Compound | $T_c$ (K) | $r_R^{Sh}$ (Å)[a] | $r_R^{P}$ (Å)[b] | $a$ (Å) | $c$ (Å) | $d$(Ni-Ni), $d$(R-C), (Å) | $d(Ni_{pl}-B_{pl})$, (Å) | $J^-$ [c] | Referenc. |
|---|---|---|---|---|---|---|---|---|---|---|
| 1 | $LaNi_2B_2C$ | 0 | 1.300 | 1.15 | 3. 794 | 9.822 | 2.6828 | 0.9743 | 11.45 | 26 |
| 2 | $PrNi_2B_2C$ | 0 | 1.266 | 1.09 | 3.7066 | 9.9999 | 2.6209 | 1.0149 | 9.38 | 6 |
| 3 | $PrNi_2B_2C$ | 0 | 1.266 | 1.09 | 3.712 | 10.036 | 2.6248 | 1.0297 | 9.54 | 26 |
| 4 | $NdNi_2B_2C$ | 0 | 1.249 | 1.08 | 3.6780 | 10.0814 | 2.601 | 1.0353 | 9.08 | 6 |
| 5 | $NdNi_2B_2C$ | 0 | 1.249 | 1.08 | 3.686 | 10.097 | 2.6064 | 1.0460 | 9.26 | 26 |
| 6 | $GdNi_2B_2C$ | 0 | 1.193 | 1.02 | 3.575 | 10.354 | 2.5279 | 1.1068 | 7.74 | 26 |
| 7 | $TbNi_2B_2C$ | 0 | 1.180 | 1.00 | 3.5536 | 10.4352 | 2.5128 | 1.1207 | 7.28 | 6 |
| 8 | $TbNi_2B_2C$ | 0 | 1.180 | 1.00 | 3.560 | 10.463 | 2.5173 | 1.1279 | 7.30 | 26 |
| 9 | $DyNi_2B_2C$ | 6 | 1.167 | 0.99 | 3.5342 | 10.4878 | 2.4991 | 1.1369 | 7.14 | 6 |
| 10 | $DyNi_2B_2C$ | 6 | 1.167 | 0.99 | 3.542 | 10.501 | 2.5046 | 1.1404 | 7.16 | 26 |
| 11 | $HoNi_2B_2C$ | 8 | 1.155 | 0.97 | 3.517 | 10.522 | 2.4869 | 1.1458 | 6.83 | 27 |
| 12 | $HoNi_2B_2C$ | 8 | 1.155 | 0.97 | 3.5177 | 10.5278 | 2.4874 | 1.1496 | 6.87 | 6 |
| 13 | $HoNi_2B_2C$ | 8 | 1.155 | 0.97 | 3.527 | 10.560 | 2.4940 | 1.1553 | 6.85 | 26 |
| 14 | $ErNi_2B_2C$ | 11 | 1.144 | 0.96 | 3.5019 | 10.5580 | 2.4762 | 1.1582 | 6.72 | 6 |
| 15 | $ErNi_2B_2C$ | 11 | 1.144 | 0.96 | 3.509 | 10.582 | 2.4812 | 1.1661 | 6.76 | 26 |
| 16 | $TmNi_2B_2C$ | 11 | 1.134 | 0.95 | 3.4866 | 10.5860 | 2.4654 | 1.1623 | 6.53 | 6 |
| 17 | $TmNi_2B_2C$ | 11 | 1.134 | 0.95 | 3.494 | 10.613 | 2.4706 | 1.1770 | 6.64 | 26 |
| 18 | $YbNi_2B_2C$ | 0 | 1.125 | 0.94 | 3.4782 | 10.607 | 2.4595 | 1.1593 | 6.30 | 6 |
| 19 | $YbNi_2B_2C$ | 0 | 1.125 | 0.94 | 3.483 | 10.633 | 2.4629 | 1.1856 | 6.54 | 25 |
| 20 | $Y_{0.96}Ni_{2.02}B_{1.96}C_{0.93}$ | 15.5 | 1.159 | 0.93 | 3.526 | 10.534 | 2.4933 | 1.1472 | 6.34 | 28 |
| 21 | $YNi_2B_2C$ | 15.5 | 1.159 | 0.93 | 3.527 | 10.536 | 2.4940 | 1.1474 | 6.34 | 29 |
| 22 | $Y_{1.02}Ni_{2.02}B_{2.02}C_{0.95}$ | 15.5 | 1.159 | 0.93 | 3.526 | 10.536 | 2.4933 | 1.1579 | 6.46 | 28 |
| 23 | $YNi_2B_2C$ | 13.5 | 1.159 | 0.93 | 3.5264 | 10.5404 | 2.4935 | 1.1447 | 6.29 | 30 |
| 24 | $YNi_2B_2C$ | 14.0 | 1.159 | 0.93 | 3.5263 | 10.5411 | 2.4935 | 1.1690 | 6.58 | 30 |
| 25 | $YNi_2B_2C$ | 14.1 | 1.159 | 0.93 | 3.5265 | 10.5417 | 2.4936 | 1.1606 | 6.47 | 30 |
| 26 | $YNi_2B_2C$ | 15.6 | 1.159 | 0.93 | 3.5257 | 10.5420 | 2.4930 | 1.1501 | 6.35 | 30 |
| 27 | $YNi_2B_2C$ | 14.0 | 1.159 | 0.93 | 3.5265 | 10.5421 | 2.4936 | 1,1691 | 6.57 | 30 |
| 28 | $YNi_2B_2C$ | 14.1 | 1.159 | 0.93 | 3.5266 | 10.5424 | 2.4937 | 1.1702 | 6.59 | 30 |
| 29 | $YNi_2B_2C$ | 14.9 | 1.159 | 0.93 | 3.5273 | 10.5426 | 2.4942 | 1.1513 | 6.36 | 30 |
| 30 | $YNi_2B_2C$ | 15.5 | 1.159 | 0.93 | 3.526 | 10.543 | 2.4932 | 1.1386 | 6.22 | 11 |
| 31 | $YNi_2B_2C$ | 15.6 | 1.159 | 0.93 | 3.524 | 10.549 | 2.492 | 1.1567 | 6.40 | 31 |
| 32 | $YNi_2B_2C$ | 15.5 | 1.159 | 0.93 | 3.533 | 10.566 | 2.4982 | 1.1496 | 6.29 | 26 |
| 33 | $LuNi_2B_2C$ | 16.6 | 1.117 | 0.93 | 3.464 | 10.631 | 2.449 | 1.1886 | 6.42 | 26 |
| 34 | $LuNi_2B_2C$ | 16.6 | 1.117 | 0.93 | 3.4639 | 10.6313 | 2.449 | 1.1918 | 6.46 | 32 |
| 35 | $LuNi_2B_2C$ | 16.6 | 1.117 | 0.93 | 3.465 | 10.633 | 2.4501 | 1.1921 | 6.46 | 29 |
| 36 | $ScNi_2B_2C$ | 13.5 | 1.010 | 0.81 | 3.37 | 10.703 | 2.383 | 1.196[f] | 5.03 | 4 |
| **37** | **$YNi_2B_2C$ [d]** | **15.5** | **1.159** | **0.93** | **3.527** | **10.543** | **2.494** | **1.155[f]** | **6.41** | **33** |
| 38 | $Y_{0.9}Lu_{0.1}Ni_2B_2C$[d] | 15.2 | 1.159[e] | 0.93 | 3.522 | 10.552 | 2.490 | 1.157[f] | 6.40 | 33 |
| 39 | $Y_{0.85}Lu_{0.15}Ni_2B_2C$[d] | 14.55 | 1.159[e] | 0.93 | 3.516 | 10.560 | 2.486 | 1.160[f] | 6.39 | 33 |
| 40 | $Y_{0.5}Lu_{0.5}Ni_2B_2C$[d] | 14.5 | 1.159[e] | 0.93 | 3.497 | 10.595 | 2.473 | 1.168[f] | 6.35 | 33 |
| 41 | $Y_{0.3}Lu_{0.7}Ni_2B_2C$[d] | 15 | 1.143[e] | 0.93 | 3.486 | 10.608 | 2.465 | 1.172[f] | 6.34 | 33 |
| 42 | $LuNi_2B_2C$[d] | 16.6 | 1.117 | 0.93 | 3.463 | 10.635 | 2.449 | 1.179[f] | 6.30 | 33 |
| **43** | **$TbNi_2B_2C$** | **0** | **1.180** | **1.00** | **3.5531** | **10.4489** | **2.5124** | **1.1253** | **7.30** | **18** |
| 44 | $Tb_{0.95}Y_{0.05}Ni_2B_2C$ | 0 | 1.180[e] | 1.00 [f] | 3.5507 | 10.4467 | 2.5107 | 1.1554 | 7.78 | 18 |
| 45 | $Tb_{0.5}Y_{0.5}$Ni2B2C | 0 | 1.180[e] | 1.00 [f] | 3.5399 | 10.4934 | 2.5031 | 1.1322 | 7.21 | 18 |
| 46 | $Tb_{0.45}Y_{0.55}Ni_2B_2C$ | 0 | 1.177[e] | 0.994 [f] | 3.5388 | 10.4978 | 2.5023 | 1.1474 | 7.33 | 18 |
| 47 | $Tb_{0.4}Y_{0.6}Ni_2B_2C$ | 2.06 | 1.175[e] | 0.987 [f] | 3.5371 | 10.5035 | 2.5011 | 1.1396 | 7.08 | 18 |
| 48 | $Tb_{0.35}Y_{0.65}Ni_2B_2C$ | 4.6 | 1.173[e] | 0.980 [f] | 3.5354 | 10.5086 | 2.4999 | 1.1108 | 6.59 | 18 |
| 49 | $YNi_2B_2C$ | 15.6 | 1.159 | 0.93 | 3.5270 | 10.5416 | 2.4950 | 1.1511 | 6.36 | 18 |
| 50 | $CeNi_2B_2C$ | 0 | 1.343[e] | 1.182 [f] | 3.637 | 10.224 | 2.573 | 1.074[f] | 12.11 | 3 |
| | 4+- 0.15 | | | | | | | | | |
| | 3+-0.85 | | | | | | | | | |

(a) $r_R^{Sh}$ - ionic crystal radius (CR, CN=8) of Shannon [34].

(b) $r_R^{P}$ - ionic radius (CN=6) of Pauling [35].

(c) $J^- = a/((c/4 - r_R^{P,eff}) - d(Ni_{pl} - B_{pl}))$

(d) Structural parameters and $T_c$, bound from the plots.

(e) $r^{eff} = S r_R \overline{(Z_R/3)}$

(f) $\boldsymbol{d(Ni_{pl} - B_{pl})}$, calculated.

where $m_n$ - contents of $R_n$-ion in a plane, $r_{R_n}$ - its radius, and $Z_{R_n}/3$ - the dimensionless factor for the registration of influence of a field created by a charge of this ion (it is equal to the ratio of a charge of ion R to a charge of yttrium ion); $S$ is a deviation factor of parameters of the R-ion, to form a plane from averaged parameters:

$$S \geq 1,\ S = \overline{r_R(Z_R/3)}/r_R(Z_R/3) \text{ or } S = r_R(Z_R/3)/\overline{r_R(Z_R/3)} \qquad (3)$$

The effective ionic radius R is equal to radius of this ion in borocarbides with one sort of tervalent rare-earth ions.

As in works [3] and [4], we have plotted the dependence of Tc from the proximate distances Ni-Ni in a plane ($T_c$(d(Ni-Ni)), Fig. 1a), Fig. 1a), from the ratio of parameters of lattice *c/a* ($T_c$(*c/a*), Fig. 1b) and from ionic radii ($r_R^{Sh}$ и $r_R^P$) of rare earth ions from Shannon system (CR, CN =8) [34] ($T_c(r_R^{Sh})$, Fig. 1c) and from the Pauling system [35] ($T_c(r_R^P)$, Fig. 1d). From calculations of all equations of curves 7 points, located outside of curve on an axis *x* and appertaining to the non-superconducting borocarbides of La, Pr, Nd, Gd and Ce, are eliminated. From of d(Ni-Ni)≤2.50 Å, *c/a*≥2.97, $r_R^{Sh,eff}$ = 1.17 Å $r_R^{P,eff}$ = 0.99 Å, which belong $DyNi_2B_2C$ и $Y_{1-x}Tb_xNi_2B_2C$ (x = 0.35 and 0.40) there are superconducting properties.

The dependencies $T_c$(d(Ni-Ni)), $T_c$(*c/a*) and $T_c(r_R^{Sh})$ (Fig. 1a, 1b and 1c) for all viewed borocarbides, except yttrium one, are close to curves set by the equations of polynomial of the second degree (approximating 93%, 92% and 96%, accordingly, Tabl. 2). In this connection the equations of curves $T_c$(d(Ni-Ni)), $T_c$(*c/a*) and $T_c(r_R^{Sh})$ were defined on parameters only of 22 compounds without yttrium borocarbides. The similar deviation for borocarbides containing yttrium from dependence between $T_c$ and metal radii of rare earths elements from the system Teatum also is observed in [4 on Fig. 2]. Nevertheless, the authors of [4] have made the contradictory conclusion about deviation from this dependence not borocarbides of yttrium, but borocarbides of four rare earths - Dy, Ho, Er and Tm, which together with borocarbides Lu and Sc lay on one curve of $\boldsymbol{T_c(r_R)}$. Such conclusion is based on earlier [3] research of $T_c$ change with ionic radii of rare earths, where separate curves for $RNi_2B_2C$ with magnetic and nonmagnetic ions R for the first time were constructed. However, we consider, that the temperatures of $T_c$ transition of $La_{0.5}Lu_{0.5}Ni_2B_2C$ (14.8 K) and $La_{0.5}Th_{0.5}Ni_2B_2C$ (3.9 K) compounds laid to the base of curve plotting for nonmagnetic R in [3], require improvement. It follows from the later works, that inappreciable concentration of large on a size impurities of rare earths results in sharp lowering $T_c$. So introduction of 10% Pr in $YNi_2B_2C$ ($\Delta r_R^P$ = 0.16 Å) reduces $T_c$ from 15.6 K up to 12.1 K [16], and of 10% La in $HoNi_2B_2C$ ($\Delta r_R^P$ = 0.18 Å) from 8.5 K to 2 K [14]. Under [4], no superconductivity is observed above 1.9 K for $Y_{0.4}La_{0.6}Ni_2B_2C$ ($\Delta r_R^P$ = 0.22 Å).

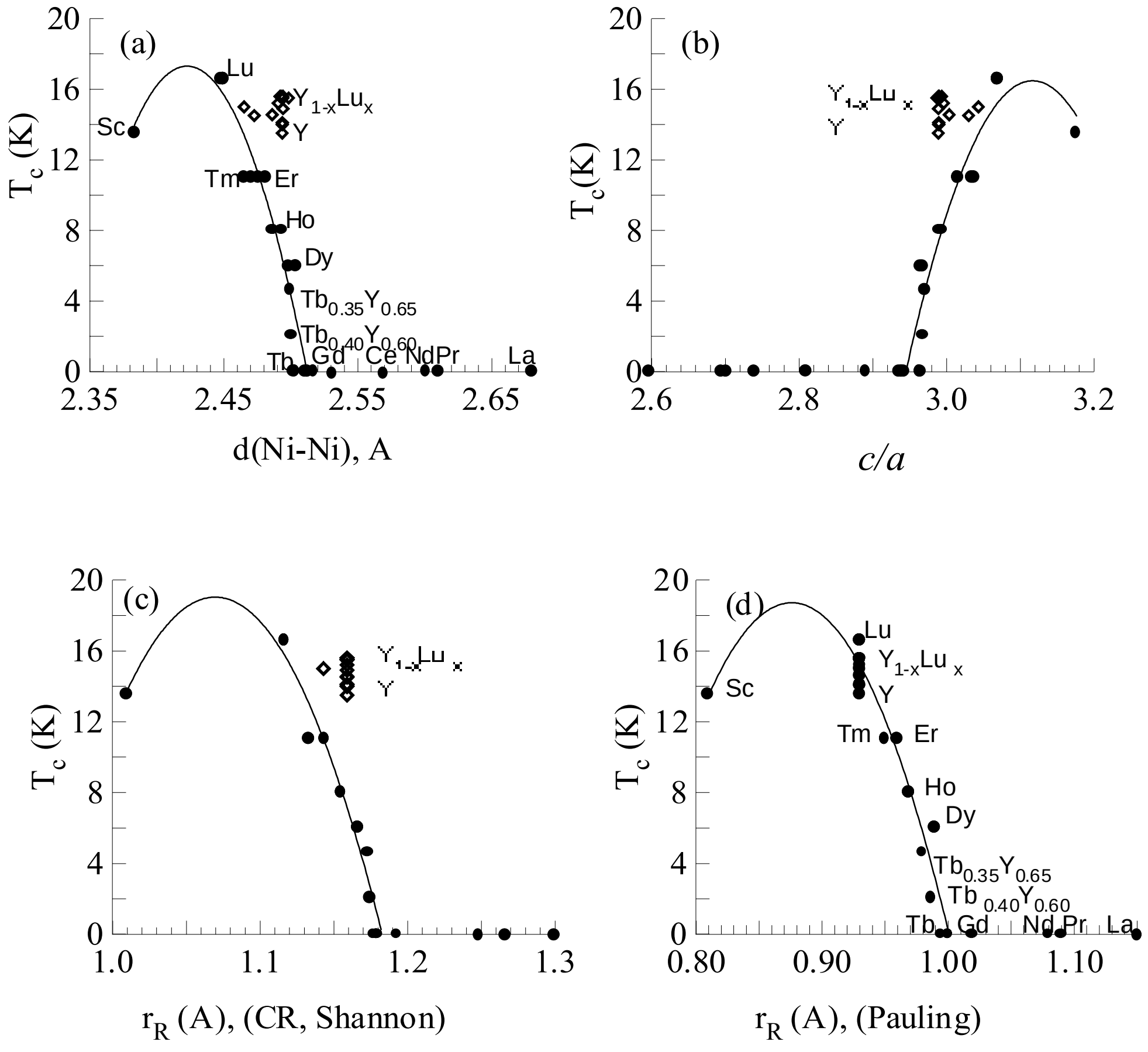

Fig.1. The variation of $T_c$ in $RNi_2B_2C$ as a function of the d(Ni-Ni) distances (a), $c/a$ ratio (b), Shannon crystal ion radii (c) and Pauling ion radii (d) of R.

Only ionic radii of rare earths from the Pauling system perfectly correlate with $T_c$ of all borocarbides irrespective of their magnetic properties (approximation to polynomial of two degree is 96 %, Tabl. 2) (Fig. 1d). Only inappreciable decrease of Lu radius could still improve correlation $T_c(r_R^{P,eff})$. We tried to find the correlation $T_c$ with other structural parameters of borocarbides, such as: parameters of lattice $a$ and $c$, interatomic distances d(Ni-Ni), d(Ni-B), d(R-B), d(R-C), d(B-C), corners B-Ni-B, distances between planes of Ni and B, R-B. All these parameters, except for lengths of Ni-B and B-C bonds, to a greater or lesser extent depend on a size of a rare-earth ion. However, any other parameters, except for ionic radii of the Pauling system, do not give correlation with $T_c$, which would approach for all R, as $T_c(r_R^{P,eff})$. Therefore, the existence of common for all R

**Table 2.** Polynomial Coefficients

| N | Function | Total points | R-squared [(a)] (%) | Polynomial coefficients | | |
|---|---|---|---|---|---|---|
| | | | | Degree 0 | Degree 1 | Degree 2 |
| 1. | $T_c$(d(Ni-Ni)) | 22[(b)] | 93 | -12636 | 10446.8 | -2156.27 |
| 2. | $T_c(c/a)$ | 22[(b)] | 92 | -5414.14 | 3485 | -559.111 |
| 3. | $T_c(r_R^{Sh,eff.})$ | 22[(b)] | 96 | -1678.55 | 3174.56 | -1484.15 |
| 4. | $T_c(r_R^{P,eff.})$ | 40 | 96 | -902.436 | 2103.52 | -1200.89 |
| 5. | $T_c r_R^{present,eff.})$ | 18[(c)] | 100 | -903.883 | 2106.63 | -1202.51 |
| 6. | $T_c(c/4 - r_R^{P,eff.})$ | 40 | 94 | -2081.87 | 2354.97 | -660.164 |
| 7. | $T_c(a/(c/4 - r_R^{P,eff.}))$ | 40 | 90 | -1015.55 | 1066.75 | -275.188 |
| 8. | $T_c(a/((c/4 - r_R^{P,eff.}) + d(Ni_{pl.} - B_{pl.})))$ | 39[(d)] | 84 | -1479 | 2562.38 | -1096.69 |
| 9. | $T_c((c/4 - r_R^{P,eff.}) - d(Ni_{pl.} - B_{pl.}))$ | 39[(d)] | 92 | -474.077 | 1637.73 | -1359.87 |
| 10. | $T_c(a/((c/4 - r_R^{P,eff.}) - d(Ni_{pl.} - B_{pl.})))$ | 39[(d)] | 94 | -245.742 | 91.104 | -7.86927 |
| 11. | $T_c(a/((c/4 - r_R^{present,eff.}) - d(Ni_{pl.} - B_{pl.})))$ | 18[(c)] | 98 | -242.636 | 90.4639 | -7.84723 |
| 12. | $\boldsymbol{d(Ni_{pl.} - B_{pl.})(c)}$ | 50 | 98 | -1.5422 | 0.255841 | - |

[(a)] Coefficient of determination (degree = 2)
[(b)] Y borocarbides was excluded from calculation.
[(c)] By calculation the $RNi_2B_2C$ with maximum $\boldsymbol{T_c}$ and reliable *z* coordinates of B atom (N 7-17, 22, 31, 33-36, 43 from Table. 1).
[(d)] Borocarbide $Tb_{0.35}Y_{0.65}Ni_2B_2C$ (N48 in Tabl. 1) was excluded from calculation.

dependence of $T_c$ from effective radii of R ions shows that a size of a rare-earth ion, its charge and dispersion of these parameters at substitutions, but not the magnetic properties of R ion, are of great important in occurrence of superconducting properties and definition of $T_c$. The relative values of rare earth radii in the Pauling system differ from other systems mainly only in radius of yttrium. It is extremely important to find the valid proofs of legitimacy of this difference to confirm existence of correlation of $T_c(r_R^{P,eff})$.

## 3. PROBLEM OF RADII OF RARE EARTHS

It is known that interpretation of structural changes and properties of compounds depends, in a great degree, on the choice of the radii system, which correctly reflects the regularity in radii changes in a number of compounds researched. The change of interatomic distances and, as a consequence, periods of lattice in $RNi_2B_2C$ (R = Sc, Y, La, Pr, Nd, Gd, Tb, Dy, Ho, Er, Tm, Yb and Lu) is linear with increase of radius R from both systems (Shannon and Pauling) for all R ions, except for Y (Fig. 2a and 2b). The abnormal behaviour of yttrium radius is the most in the Pauling system. According to correlations $a(r_R^{Sh})$, $a(r_R^P)$, $c(r_R^{Sh})$ and $c(r_R^P)$ the yttrium radius in the Shannon system is a little overestimated, and in the Pauling system strongly diminished. The size of yttrium ion should be intermediate between the sizes of Ho and Er ions.

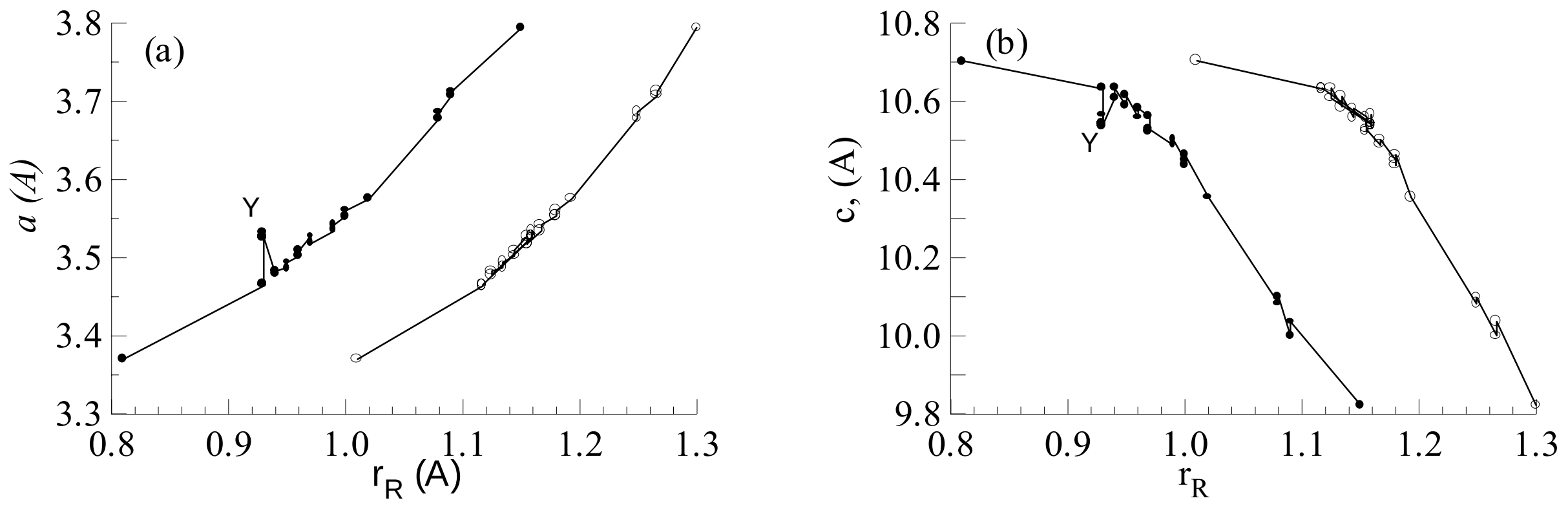

**Fig. 2**. The variation of the lattice parameters *a* (a) and *c* (b) in $RNi_2B_2C$ as function of Shannon crystal ion radii (solid symbols (•)) and Pauling ion radii (open symbols (◦)).

In most cases the radii of atoms and ions are calculated from interatomic distances. However, interatomic distances depend on great number of factors and not always adequately reflect a size of an ion. Such fact takes place in a number of viewed three-valent rare-earth ions. Under [36] about 10% of lanthanoid contraction is relativistic. The relativistic contraction of orbitals and relativistic contraction of bond lengths are two parallel, but largely independent effects. As the radii Y and Lu in the Pauling system are equal, we can estimate what would be bond lengths of Lu-B and Lu-C in absence of relativistic contraction. Let's assume, that 10% from cutting of bond lengths of R-B (Δ=0.209 Å) and R-C (Δ=0.234 Å) at transition from La (d(La-B)=3.065 Å, d(La-C)=2.683 Å [26 ]) to Lu (d(Lu-B)=2.856 Å, d(Lu-C)=2.449 Å [26]) concern relativistic bond-length contraction. Then, without relativistic contraction of bond lengths, caused by nuclear charge increase ((Z) on 14, the bond lengths in в $LuNi_2B_2C$ d(Lu-B) and d(Lu-C) would be equal 2.877 Å and 2.472 Å, accordingly, and *a* ( $a = \sqrt{2}$ d(R-C); d(R-C)=d(Ni-Ni)) would be equal 3.496 Å instead of 3.464 Å, i.e. their values are close to ones for Er borocarbide. However, the lengths of bond d(Lu-B) and d(Lu-C) increase for 0.016 Å on 0.019 Å, accordingly if to assume, that in absence of relativistic contraction of bond at transition from Y to La the increase of length of bond d(Lu-B) and d(Lu-C) would be on 10% more.

In result, at absence of relativistic contraction the bond length d(R-B) and d(R-C) and parameter *a* in lanthanum borocarbide would be equal 3.081 Å, 2.702 Å and 3.821 Å, in lutecium borocarbide of 2.893 Å, 2.491

Å and 3.527 Å, accordingly, that is close to experimental values for yttrium borocarbide (2.902 Å, 2.493 Å and 3.526 Å [30]. Therefore, Y radius is close to Lu one, as follows from the system of ionic Pauling radii [35], instead of to Ho radius on the Shannon (CR, CN = 8) system [34].

**Table 3** Ionic Radii for rare earth elements

| Element | $r^{Sh}$ (Shannon) [(a)] | $r^{Sh}$ (present) [(b)] | $r^{P}$ (Pauling) [(a)] | $r^{P}$ (present) [(b)] |
|---|---|---|---|---|
| Sc | 1.010 | 1.008 | 0.81 | 0.810 |
| Y | 1.159 | 1.118 | 0.93 | 0.928 |
| Tb | 1.180 | 1.183 | 1.00 | 1.001 |
| Dy | 1.167 | 1.163 | 0.99 | 0.979 |
| Ho | 1.155 | 1.156 | 0.97 | 0.970 |
| Er | 1.144 | 1.143 | 0.96 | 0.956 |
| Tm | 1.134 | 1.143 | 0.95 | 0.956 |
| Lu | 1.117 | 1.110 | 0.93 | 0.918 |

*(a)* Ionic radii of Shannon (CR, CN=8) [34] and of Pauling (CN=6) [35].
*(b)* $r^{Sh}$ and $r^{P}$ from r *vs* $T_c$ plots.

This problem was already appeared at research of correlations in HTSC cuprates. However, we have not pointed at it attention, as viewed various systems with Ba, Sr, Ca, Y and Ln cations, having the difference both in sizes and in charges. In this case $RNi_2B_2C$ borocarbides with small differences in size of R cations and the same charge any alteration of a relative size of ions are deciding. We calculated the radii of rare earth ions for the Shannon (CR, CN = 8) and Pauling systems (Tabl. 3) from the dependences $\boldsymbol{T_c(r_R^{Sh,eff})}$ and $\boldsymbol{T_c(r_R^{P,eff})}$ found by us. The ionic radius Y appears only on 0.01 Å more than Lu one. It explains inappreciable lowering Tc with x increase in $Lu_{1-x}Y_xNi_2B_2C$ [33].

Using radii, precised by us, for the Pauling system, we have calculated parabolic correlation $\boldsymbol{T_c(r_R^{present,eff})}$ (Tabl. 2, function (5)). Maximal $T_c$ values for borocarbides Ni, calculated on the correlation are 18, 75 K at optimal value $\boldsymbol{r_o^{eff}} = 0.876$. Then, on function (5) (Tabl.2), we have determined the changes $T_c$ in $R_{1-x}R'_xNi_2B_2C$ with increasing x for systems with $\boldsymbol{r_R^{eff}} > \boldsymbol{r_o^{eff}}$, such as: $Lu_{1-x}R'_xNi_2B_2C$ (R' = Y, Ho, La) (I), $Y_{1-x}R'_xNi_2B_2C$ (R' =Ho, Pr, Tb, La) (II) and $Ho_{1-x}R'_xNi_2B_2C$ (R' = Dy, La) (III), and also for systems with $\boldsymbol{r_R^{eff}} < \boldsymbol{r_o^{eff}}$, such as: $Sc_{1-x}R'_xNi_2B_2C$ (R' = Lu, Y, Ho, La), $Sr_{1-x}R'_xNi_2B_2C$ (R' = Ho, La), $Ca_{1-x}R'_xNi_2B_2C$ (R' =Sc, La) (Tabl. 4). It was found that parabolic dependence $\boldsymbol{T_c(r_R^{present,eff})}$ adequately to the experimental data [14-17, 19, 33] estimate a trend of changing $T_c$ of I, II and III systems with increasing x from 0 to 0.5. At substitution R on R' with smaller value $r^{eff}$, than for R, the $T_c$ of borocarbides, located in the left parabola part, decreases and in right, one opposite - increases, passing through a point of maximum by a curve, and then decreases. At substitution R on R' with larger $r^{eff}$, the $T_c$ of borocarbides, located in the left part of a curve, raises, passing through maximum, and in right part, opposite, decreases.

**Table 4**. Calculated $r_R^{present,eff.}$ and $T_c$ in $R_{1-x}R'_xNi_2B_2C$.

| $R_{1-x}R'_x$ | $r_R^{present,eff.}$ | $T_c$, K | $R_{1-x}R'_x$ | $r_R^{present,eff.}$ | $T_c$, K |
|---|---|---|---|---|---|
| Sc | 0.810 | 13,5 | Lu | 0.918 | 16.6 |
| $Sc_{0.80}Lu_{0.20}$ | 0.854 | 18.2 | $Lu_{0.90}Y_{0.10}$ | 0.920 | 16.4 |
| $Sc_{0.75}Lu_{0.25}$ | 0.865 | 18.6 | $Lu_{0.60}Y_{0.40}$ | 0.926 | 15.7 |
| $Sc_{0.70}Lu_{0.30}$ | 0.876 | 18.75 | $Lu_{0.50}Y_{0.50}$ | 0.928 | 15.6 |
| $Sc_{0.65}Lu_{0.35}$ | 0.887 | 18.6 | $Lu_{0.10}Y_{0.90}$ | 0.928 | 15.6 |
| $Sc_{0.50}Lu_{0.50}$ | 0.918 | 16.6 | $Lu_{0.90}Ho_{0.10}$ | 0.928 | 15.6 |
| $Sc_{0.90}Y_{0.10}$ | 0.834 | 16.6 | $Lu_{0.80}Ho_{0.20}$ | 0.939 | 14.0 |
| $Sc_{0.75}Y_{0.25}$ | 0.870 | 18.7 | $Lu_{0.70}Ho_{0.30}$ | 0.949 | 12.3 |
| $Sc_{0.60}Y_{0.40}$ | 0.907 | 17.6 | $Lu_{0.60}Ho_{0.40}$ | 0.960 | 10.1 |
| $Sc_{0.50}Y_{0.50}$ | 0.928 | 15.6 | $Lu_{0.55}Ho_{0.45}$ | 0.965 | 9.2 |
| $Sc_{0.80}Ho_{0.20}$ | 0.875 | 18,7 | $Lu_{0.50}Ho_{0.50}$ | 0.970 | 8.0 |
| $Sc_{0.70}Ho_{0.30}$ | 0.909 | 17.4 | $Lu_{0.90}La_{0.10}$ | 0.965 | 9.2 |
| $Sc_{0.60}Ho_{0.40}$ | 0.943 | 13.3 | $Lu_{0.85}La_{0.15}$ | 0.989 | 3.6 |
| $Sc_{0.90}La_{0.10}$ | 0.844 | 17.5 | $Lu_{0.80}La_{0.20}$ | 1.013 | Non-Super. |
| $Sc_{0.80}La_{0.20}$ | 0.952 | 11.8 | Y | 0.928 | 15.6 |
| $Sc_{0.75}La_{0.22}$ | 0.989 | 3.4 | $Y_{0.90}Ho_{0.10}$ | 0.936 | 14.4 |
| $Sc_{0.70}La_{0.30}$ | 1.026 | Non-Super. | $Y_{0.80}Ho_{0.20}$ | 0.945 | 13.0 |
| Sr | 0.753 | 0.6 | $Y_{0.70}Ho_{0.30}$ | 0.953 | 11.6 |
| $Sr_{0.80}Ho_{0.20}$ | 0.842 | 17.4 | $Y_{0.60}Ho_{0.40}$ | 0.962 | 9.8 |
| $Sr_{0.75}Ho_{0.25}$ | 0.865 | 18.6 | $Y_{0.50}Ho_{0.50}$ | 0.970 | 8.0 |
| $Sr_{0.70}Ho_{0.30}$ | 0.889 | 18.5 | $Y_{0.90}Tb_{0.10}$ | 0.943 | 13.3 |
| $Sr_{0.60}Ho_{0.40}$ | 0.937 | 14.3 | $Y_{0.65}Tb_{0.35}$ | 0.980 | 6.2 |
| $Sr_{0.55}Ho_{0.45}$ | 0.961 | 10.1 | $Y_{0.62}Tb_{0.38}$ | 0.984 | 4.7 |
| $Sr_{0.50}Ho_{0.50}$ | 0.970 | 8.0 | $Y_{0.60}Tb_{0.40}$ | 0.987 | 3.9 |
| $Sr_{0.90}La_{0.10}$ | 0.834 | 16.6 | $Y_{0.57}Tb_{0.43}$ | 0.992 | 2.5 |
| $Sr_{0.85}La_{0.15}$ | 0.877 | 18.75 | $Y_{0.48}Tb_{0.52}$ | 0.999 | 0.5 |
| $Sr_{0.80}La_{0.20}$ | 0.920 | 16.4 | $Y_{0.50}Tb_{0.50}$ | 1.01 | 0 |
| $Sr_{0.75}La_{0.25}$ | 0.965 | 9.2 | $Y_{0.94}Pr_{0.06}$ | 0.948 | 12.5 |
| $Sr_{0.71}La_{0.29}$ | 1.00 | 0 | $Y_{0.90}Pr_{0.10}$ | 0.961 | 10.0 |
| Ca | 0.66 | Non-Super. | $Y_{0.80}Pr_{0.20}$ | 0.994 | 2.0 |
| $Ca_{0.70}Sc_{0.30}$ | 0.753 | 0.6 | $Y_{0.75}Pr_{0.25}$ | 1.011 | Non-Super. |
| $Ca_{0.65}Sc_{0.35}$ | 0.769 | 4.7 | $Y_{0.90}La_{0.10}$ | 0.973 | 7.4 |
| $Ca_{0.60}Sc_{0.40}$ | 0.785 | 8.8 | $Y_{0.85}La_{0.15}$ | 0.996 | 1.4 |
| $Ca_{0.50}Sc_{0.50}$ | 0.810 | 13.5 | $Y_{0.80}La_{0.20}$ | 1.018 | Non-Super. |
| $Ca_{0.90}La_{0.10}$ | 0.762 | 3.1 | Ho | 0.970 | 8.0 |
| $Ca_{0.80}La_{0.20}$ | 0.870 | 18.7 | $Ho_{0.90}Dy_{0.10}$ | 0.972 | 7.7 |
| $Ca_{0.70}La_{0.30}$ | 0.987 | 3.9 | $Ho_{0.70}Dy_{0.30}$ | 0.975 | 7.0 |
| $Ca_{0.50}La_{0.50}$ | 1.15 | Non-Super | $Ho_{0.50}Dy_{0.50}$ | 0.979 | 6.0 |
| | | | $Ho_{0.95}La_{0.05}$ | 0.988 | 3.5 |
| | | | $Ho_{0.94}La_{0.06}$ | 0.992 | 2.5 |
| | | | $Ho_{0.90}La_{0.10}$ | 1.006 | Non-Super. |

Thus, we have shown, that:

- The deviations of lattice periods of Y borocarbide from dependence $a(r_R)$ and $c(r_R)$ (Fig. 2) are quite natural and are explained by inappreciable relativistic contraction of bond for yttrium and increase of bond contractions with increase of nuclear charge in a number of lanthanides;
- On the basis of dependence of temperature of transition in a superconducting state for nickel borocarbides $RNi_2B_2C$ from a size of a rare-earth ion ($T_c(r_R)$) it is possible to improve radii of a number of rare earths (Tabl. 3);
- The Pauling system of ionic radii most precisely reflects relative sizes of rare earths ions.

## 4. ON SIMILAR NATURE OF CORRELATION BETWEEN $T_c$ AND CRYSTAL CHEMICAL PARAMETERS IN LAYERED SUPERCONDUCTORS: HIGH-$T_c$ CUPRATES, DIBORIDES AND INTERMETALLIC BOROCARBIDES

In layered quasi-two-dimensional systems, such as borocarbides of nickel, high-$T_c$ cuprates and diborides it is possible to select similar elementary structural fragments - sandwiches I/S/I, where the internal lay S contains charge carriers capable to carry, and the external lays I are dielectric ones. In structures RNi2B2C the plane square nets formed by Ni atoms. (Fig. 3a). Half of cells of these nets is centered by R atoms from above and B atoms from below, and other half on the contrary. As a result, along axes *a* and *b* the strips of Ni cells, centered by R atoms from above, with strips of cells centered by R atoms from below alternate. If we select in a plane half of cells centered identicaly, we receive sandwiches RB(Ni), similar sandwiches in high-$T_c$ cuprates and diborides A2(Cu) (Fig. 3b) and A2(B2) (Fig. 3c) (A-positive charged ion). The internal plane of sandwiches RB(Ni), with charge carriers, is formed by square nets from Ni atoms (side of a quadrate is equal $a\sqrt{2}/2$). One of external planes of a sandwich is made by square nets from rare-earth atoms R (side of a quadrate is equal to tetragonal lattice parameter *a*), and another by the same nets from B atoms.

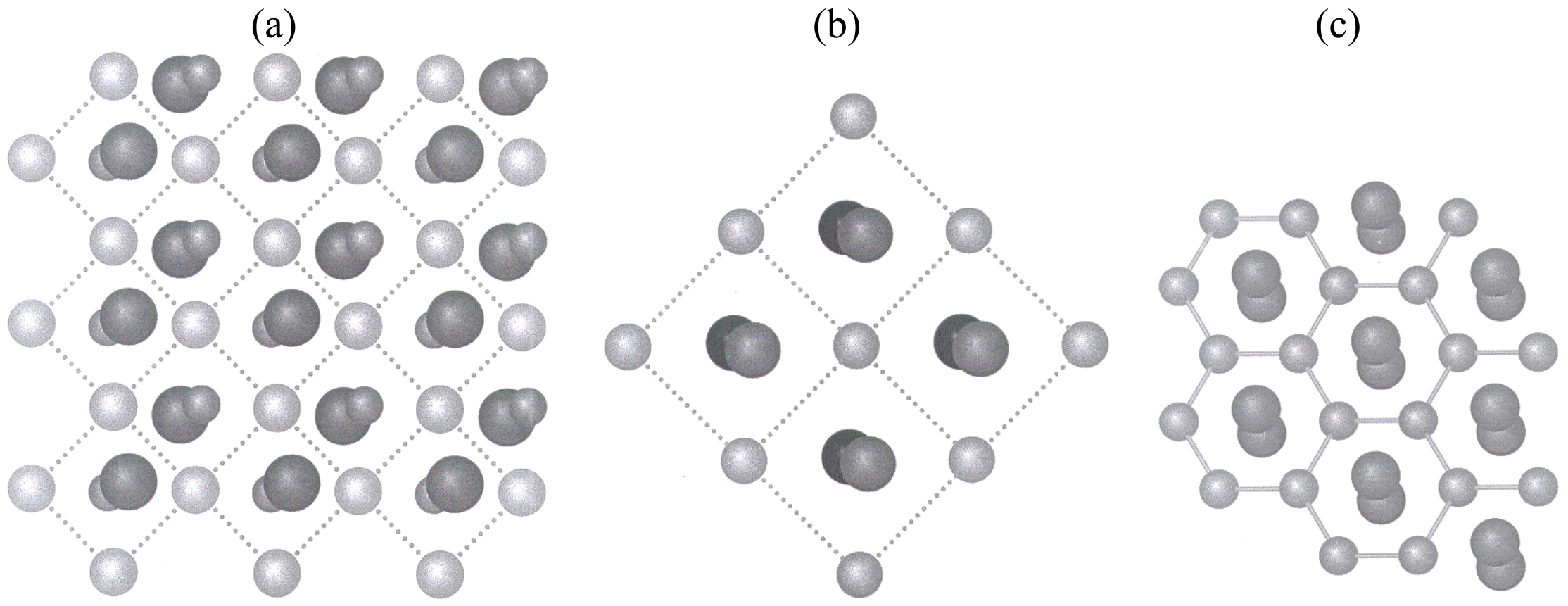

**Fig. 3**. Structural fragments – sandwiches: RB(Ni) in $RNi_2B_2C$ (R – large black ball, B – small black ball, Ni – light ball) (a), $A_2$(Cu) in HTSC cuprates (A – large ball, Cu – small ball) (b) and $A_2(B_2)$ in diborides (A-large ball, B small ball) (c).

The sensitivity of superconductivity to a size and charge of ions in external planes of a sandwich and to difference in these values by substitution, to distances between internal and external planes of a sandwich, and also to distances between atoms in an internal plane specifies the responsibility of these elementary fragments for occurrence of superconductivity. We guess, that transfer of carriers at occurrence of superconductivity in borocarbides is carried out between Ni atoms on a line of location of centering atoms R and B along the cell strips centered by an identically (along [100] and [010] directions), but not between Ni atoms, bound by short Ni-Ni bonds, i.e. as well as in high-$T_c$ cuprates [24] along diagonals of quadrates (Fig. 3a, 3b). In [23-25] we have shown that in superconducting cuprates and diborides there are parabolic correlations of the $T_c$ with a relation of crystal chemical parameters of sandwiches. Similar correlations we have detected and in borocarbides. It appears that Tc borocarbides well correlates with effective distance $D_1$ from a plane of Ni up to a plane R (approximating $T_c(D_1)$ 94 %, function N6 in Tabl. 2):

$$D_1 = c/4 - r_R^{P,eff} \tag{4}$$

where $r_R^{P,eff}$ by formula (1), and also with a relation ($J_1$) of distance between Ni atoms, located on the ends of diagonals of square cell (parameter $a$) to effective distance $D_1$ (approximating 90 %, function N7 in Tabl. 2):

$$J_1 = a/D_1 \tag{5}$$

where $D_1$ by formula (4).

However, unlice high-$T_c$ cuprates and diborides in Ni borocarbides there is not correlation between $T_c$ and ratio ($J$) of Ni-Ni in a plane to the total of effective distances ($D_1$+$D_2$) from a Ni plane up to two adjacent planes of atoms R ($D_1$) and atoms B ($D_2$=d($Ni_{plane}$-$B_{plane}$)):

$$J = a/((c/4 - r_R^{P,eff}) + d(Ni_{pl} - B_{pl})) \tag{6}$$

In this case approximating $T_c(J)$ makes only 84 % (function N8 in Tabl. 2) and to improve it fails neither by $\boldsymbol{r_R^{present,eff}}$. instead of $r_R^{P,eff}$, nor exepting from calculation the data for borocarbides with random errors in definition of coordinate $z$ of B atom. From here follows, that distance $Ni_{plane}$-$B_{plane}$ has influence on value $T_c$, but opposite to effect of change of R radius, as by decrease of radius the distance raises (Fig. 5a) because of impossibility to change length of strong covalent bond Ni-B [6].

Indeed, $T_c$ correlates not with the total of effective distances ($D_1$+$D_2$), and with their difference (approximating 92 %, function N9 in Tabl. 2). More precise dependence is between $T_c$ and ratio of Ni-Ni distances to a difference of distances ($D_1$-$D_2$) (approximating 94 %, function N10 in Tabl. 2, Fig. 4b):

$$J^- = a/((c/4 - r_R^{P,eff}) - d(Ni_{pl} - B_{pl})) \tag{7}$$

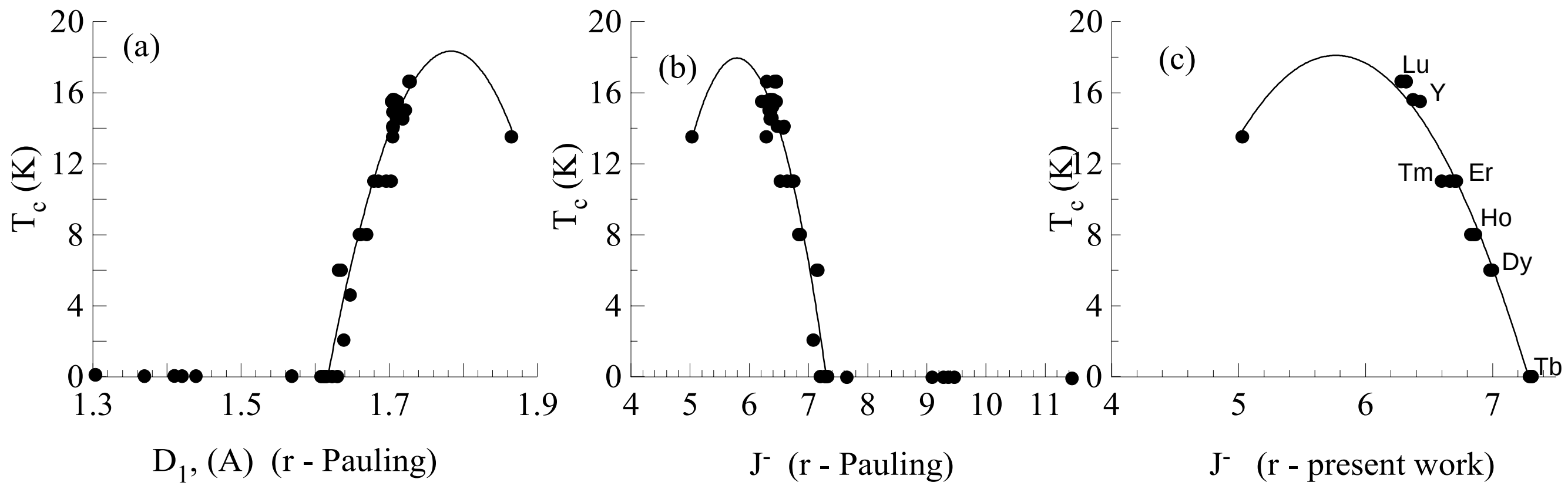

**Fig. 4.** $T_c$ as function $D_1$ (a) and $J^-$ (b), calculated by $r_R$ from the Pauling system, and calculated by $r_R$ from this work (c).

The approximating of the function $T_c(J^-)$ can be improved up to 98 %, if instead of $\boldsymbol{r_R^{P,eff.}}$ to take $\boldsymbol{r_R^{present,eff.}}$ and to eliminate from calculation the borocarbides with distances d($Ni_{pl}$-$B_{pl}$), which deviate from dependences with radii R and lattice parameters *a* and *c* (function N12, Tabl. 2, Fig. 5).

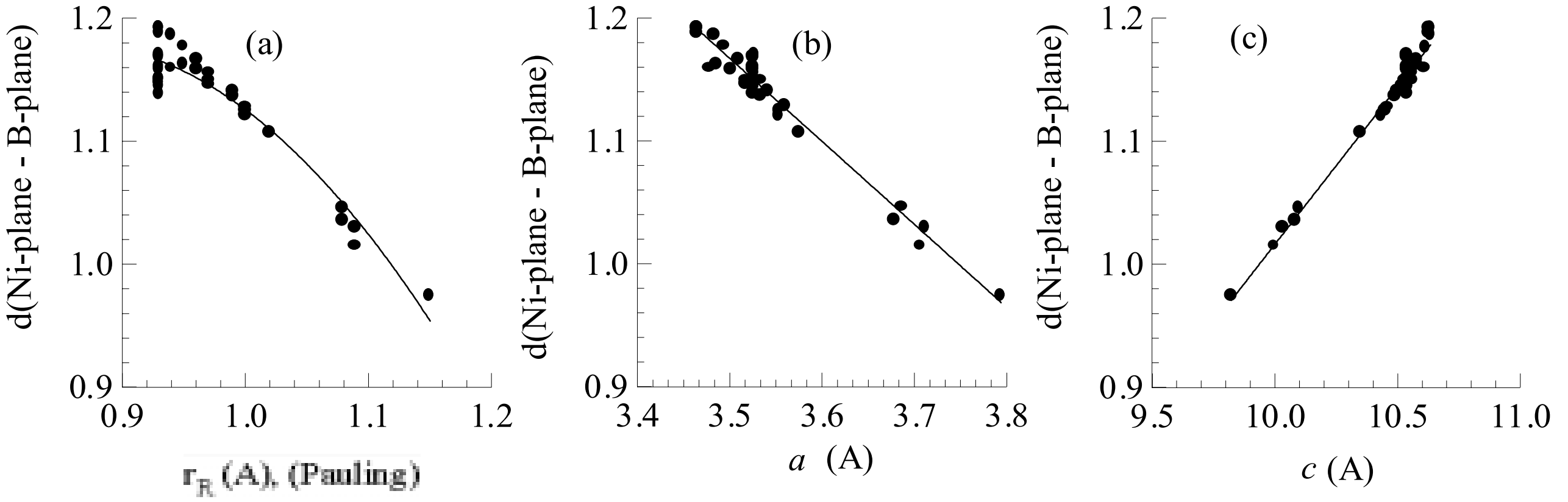

**Fig. 5.** The variation of the distance between the planes of Ni and B in RNi2B2C as a function of Pauling ion radii (a), lattice parameters *a* (b) and *c* (c).

The argument $J^-$ of function $T_c$ is defined by lattice parameters *a* and *c* and distance d($Ni_{pl}$-$B_{pl}$), which in turn (at standard atmosphere pressure) are functions from radius R (Fig. 2a, 2b and 5a), and the value $\boldsymbol{r_R^{P,eff.}}$ is defined by a size and charge R; therefore

$$T_c(J^-) = T_c(r_R^{eff.}), \tag{8}$$

and for borocarbides with one sort of a tervalent rare-earth atoms, by $\boldsymbol{r_R^{eff.}} = \boldsymbol{r_R}$, $\boldsymbol{T_c(J^-) = T_c(r_R)}$. Let's mark, that the inappreciable deterioration of approximating $\boldsymbol{T_c(J^-)}$ in comparison with $\boldsymbol{T_c(r_R^{eff.})}$ is connected, apparently, with experimental errors by definition of parameters *a* and *c*, coordinate *z* of B atom, and also composition of compounds. The maximal $T_c$ of nickel borocarbides, according to the function $\boldsymbol{T_c(J^-)}$ (function N12, Tabl. 2) is equal 18,1 K at $\boldsymbol{J_o}$ = 5.8.

The sensitivity $T_c$ to distance d($Ni_{pl}$-$B_{pl}$) is determined by correlation of of this distance with the corner between Ni atom and B atoms, located in one plane. In this case the distance d($Ni_{pl}$-$B_{pl}$) is entered, as the characteristic of width of the channel of conduction electrons transfer (Fig. 6). The channel is wider, when a distance d($Ni_{pl}$-$B_{pl}$) is less and accordingly a corner B-Ni-B is more.

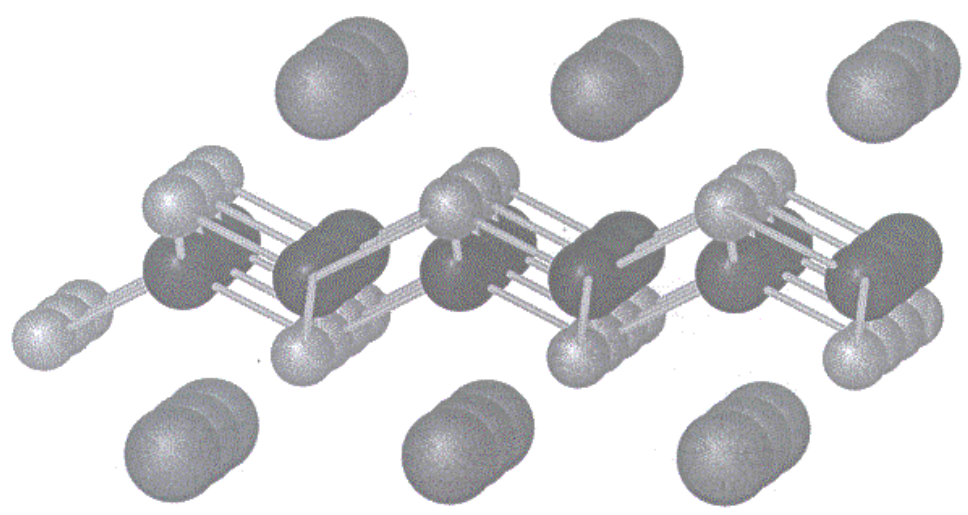

**Fig. 6.** Ni-B bonds in $RNi_2B_2C$, restricting the width of carry channels along the [100] and [010] directions conduction electrons.

Binding electronic density between Ni and B atoms, apparently, suppresses carry of charge on line of its lacation.

Under pressure $T_c$ of borocarbides with $\boldsymbol{J}^{-} > \boldsymbol{J}_{\boldsymbol{o}}^{-}$ should decrease, if the squeezing of the parameter *c* anticipates the one of the parameter *a*. It follows from dependence $\boldsymbol{T}_{\boldsymbol{c}}(\boldsymbol{J}^{-})$ (function N7 and N12, Tabl. 2, Fig. 4d and 4c) and dependences d($Ni_{pl}$-$B_{pl}$) from lattice parameters of *a* (Fig. 5b) and *c* (Fig. 5c). However, the change of parameters *a* and *c* is interdependent and is regulated by the length of covalent Ni-B bond. As a result the distance between planes of atoms Ni and B under pressure can uncertainly vary in narrow limits, as at squeezing the *c* parameter it is shortened, and at squeezing the *a* parameter, it is incremented (Fig. 5 b and c). According to dependence $\boldsymbol{T}_{\boldsymbol{c}}(\boldsymbol{J}^{-})$, such uncertainty influences on change of borocarbides $T_c$ by pressure. Really, experimental researches of influence of pressure on the superconducting state of RNi2B2C (R=Y, Ho, Er and Tm) compounds installed in [37] inappreciable decreasing $T_c$ with pressure for R=Y, Ho, Er and Tm, and in [38] only for R=Ho and Tm, whereas for R=Y and Er, opposite, $T_c$ increases with pressure.

Thus, in spite of perfectly correlates of $T_c$ with effective radii ($\boldsymbol{r}_{\boldsymbol{R}}^{\boldsymbol{P,eff}}$) of R for RNi2B2C compounds its value defines not the width of R lay, that has only one atom, but an effective charge space carrying between internal and outside planes of sandwich, calculated in view of a charge and jaggings of a surface of outside planes at R substitutions and presence of bonding electrons, and also the distances between Ni atoms in a plane. It is confirmed by change of borocarbides $T_c$ under pressure. As all numbered parameters, as was shown above, depend on a size R, for an estimation of the $T_c$ value in borocarbides both functions $\boldsymbol{T}_{\boldsymbol{c}}(\boldsymbol{J}^{-})$ and $\boldsymbol{T}_{\boldsymbol{c}}(\boldsymbol{r}_{\boldsymbol{R}}^{\boldsymbol{eff}})$ are applied.

Low values $T_c$ in borocarbides in comparison with high-$T_c$ cuprates and $MgB_2$ depend, apparently, on two reasons: (1) complete suppression of charge carry between Ni and B planes and partial suppression of carry between Ni and R planes by electrons of covalent bond Ni-B; (2) feeble focusing of charge carriers - electrons to a trajectory of a motion by a plane of atoms R, charged positively. Probably, for reaching high Tc external lays of a sandwich should have a like sign of a charge with carriers, as their function is the focusing of carriers to a trajectory of carry. Confirmation of this hypothesis can be reaching higher Tc (89 K) at optimal hole doping in comparison with electronic one (34 K) of superconducting film $CaCuO_2$ [39].

## 5. CONCLUSIONS

In borocarbides $RNi_2B_2C$ the sandwiches RB(Ni), similar existing in superconducting layered cuprates and diborides are allocated. The correlations of $T_c$ with a relation of crystal chemical parameters of these sandwiches also are described by similar parabolic correlation, common for magnetic and nonmagnetic R. According to this correlation, $T_c$ depends on such critical crystal chemical parameters as: radius and charge of R; difference in these values at substitutions of R; distances between internal and external planes of a sandwich; distance between atoms Ni in an internal plane. For borocarbides these parameters are depending on a size or size and charge of R and, therefore, $T_c$ correlates and with effective radius of R, or in non-substituted borocarbides of tervalent rare earths with ionic radii of R, but only from the Pauling system. We have analyzed of Shannon and Pauling ionic radii systems and have shown, that only Pauling system adequately estimates R sizes, as it takes into account intensification of relativistic contraction of bond lengths with increase of nuclear charge of R atoms. On the basis of the function $\boldsymbol{T_c(r_R)}$ the values of radii of rare earths are improved. The correlation found adequate to the experimental data estimates the tendency of $T_c$ change borocarbides at various R substitutions and with pressure rise.

It was shown that along the [100] and [010] directions in borocarbides, apparently, a superconducting transfer of carrying electrons takes place. In the later work [40] it is proved, that point nodes of gap function of $YNi_2B_2C$ are located along the same directions. We suppose, that the low values of $T_c$ are stipulated by suppression of transfer of carrying electrons by bonding electrons located between planes of Ni and B and feeble focusing of carrying electrons to a line of a motion by a plane of R atoms because of unlike sign of charges of carriers and a plane.

Thus, the existence in borocarbides $RNi_2B_2C$ common for magnetic and nonmagnetic R of correlation $T_c$ with a relation of crystal chemical parameters proves possibility of parallel coexisting of a superconductivity and magnetism without influence of magnetism on value $T_c$ in quasi-two-dimensional systems containing sandwiches I/S/I, where in internal lay S has charge carriers capable to superconducting transfer, and the external lays I with width of lay in one atom are dielectric ones.

## ACKNOWLEDGMENTS

This work was supported by the Russian Foundation for Basic research under grant 00-03-32486.